\documentclass[preprint2,a4paper]{aastex}
\usepackage{epstopdf}
\usepackage{emulateapj5}
\usepackage{onecolfloat5}

\newcommand{\be}{\begin{equation}}
\newcommand{\ee}{\end{equation}}

\input{coalsack.defs}

\slugcomment{To appear in The Astrophysicsl Journal}

\shorttitle{On the Magnetic Field in the Vicinity of the Southern Coalsack}
\shortauthors{Bhat \& Andersson}

\begin{document}


\twocolumn[%%% Begin front material
\title{On the magnetic field through the Upper Centaurs-Lupus Super bubble, in the vicinity of  the Southern Coalsack}

\author{N. D. Ramesh Bhat}
\affil{Centre for Astrophysics \& Supercomputing, Swinburne University of Technology, Hawthorn, Victoria 3122, Australia}  
\vskip 0.25cm
\author{B-G Andersson}
\affil{SOFIA Science Center/USRA, NASA Ames Research Center, M.S. N211-3, Moffett Field, CA 94035, USA}


\begin{abstract}
The Southern Coalsack is located in the interior of the Upper Centaurus-Lupus (UCL) super bubble and shows many traits that point to a much more energetic environment than might be expected from a dark, starless molecular cloud.  A hot, X-ray emitting, envelope surrounds the cloud, it has a very strong internal magnetic field and its darkest core seems to be on astronomical time scales ``just about" to start forming stars.  In order to probe the magnetic environment of the cloud and to compare with the optical/near infrared polarimetry-based field estimates for the cloud, we have acquired Faraday Rotation measurements towards the pulsar PSR J1210$-$6550, probing the magnetic field in the vicinity of the cloud, and a comparison target, PSR J1435$-$5954, at a similar line of sight distance but several degrees from the cloud.  Both lines of sight hence primarily probe the UCL super bubble.  The earlier estimates of the magnetic field inside the Coalsack, using the Chandrasekhar-Fermi method on optical and near-infrared polarimetry, yield B$_\perp$ = 64--93~$\mu$G.   However, even though PSR J1210$-$6550 is located only $\sim$30 arc minutes from the (CO) edge of the cloud, the measured field strength is only B$_\parallel$ = 1.1$\pm$0.2 $\mu$G.  While thus yielding a very high field contrast to the cloud we argue that this might be understood as due to the effects on the cloud by the super bubble.
\end{abstract}

]%%% End front material

\section{Introduction} \label{s:intro}

The Southern Coalsack provides one of the most spectacular sights on the southern night sky -- at least for any interstellar medium astronomer -- with its very dark nebulosity contrasted by the Southern Cross and the Jewell Box cluster (NGC~4755).   Beyond its naked eye attractiveness, it is a well-studied dark cloud, which, due to its relative vicinity (high spatial resolution) and location in the Galactic plane (abundance of background stars suitable for optical and UV spectroscopy), provides one of the best test cases for the study of star-less molecular clouds (for a thorough introduction to the cloud see the review by \citet{nyman2008}).  The cloud has been mapped in $^{12}$CO by \citet{nyman1989}, who found a total mass of $\sim$3500~M$\odot$, and in $^{13}$CO by \citet{kato1999}.   The distance to the cloud ($d$) has been discussed by a number of authors, including \citet{seidensticker1989b} who performed an extensive spectroscopic and photometric survey of field stars in the region and find that the cloud is made up of two components, one at $d\sim$188~pc and the other at $d\sim$243~pc.   Later authors, generally, do not support the dual components, but do find a cloud distance in broad agreement with these authors (see e.g. \citet{franco1989, straizys1994,knude1998}).  The study by \citet{seidensticker1989b} had the additional advantage of mapping out the extinction well beyond the cloud and they conclude that the space beyond the Coalsack is relatively free of additional extinction until the Carina Arm is reached at $\sim$1.3~kpc.  The Coalsack is located at \textit{l}$\sim$300$^\circ$.  With a nominal spiral arm pitch angle if $-11^\circ$ \citep{vallee2005} the relative angle between the line of sight and the undisturbed magnetic field direction is of the order of 20--30$^\circ$.

\citet{crawford1991} pointed out that, based on the H~I observations by \citet{degeus1992}, the cloud is located inside the Upper Centaurus-Lupus (UCL) super bubble.  \citet{degeus1992} estimates that the super bubble has a radius of R=110$\pm$10~pc, an estimated age about 11~Myr, an estimated total energy input of $\sim$0.9$\times$10$^{51}$ ergs, and depending on the adopted distance to the Coalsack, the shell would have overtaken the cloud  about 2--5 Myr ago.

\citet{bga2004} found that the extended envelope of the cloud contained the high-ionization state O~VI ion, characteristic of a gas temperature of $\sim$300,000~K \citep{shapiro1976}.  Because O~V has an ionization potential of $\sim$113~eV, O~VI is generally expected to be produced by collisional ionization in the ISM.  They also found that the cloud envelope could be seen in soft X-rays from the ROSAT PSPC observations.  Based on these observations \citet{bga2004} proposed that the hot cloud envelope was due to the cloud's envelopment by the UCL super bubble.  \citet{duncan1995} noted the existence of a 2.4 GHz continuum emission arc around the cloud, which they designated G303.5+0, and the data from \citet{duncan1997} show that this emission is polarized (see Figure 2).   \citet{walker1998} identified this radio feature with an  H$\alpha$ shell that they named ``The Coalsack Loop".

\begin{figure*}
\epsscale{1.50}
\plotone{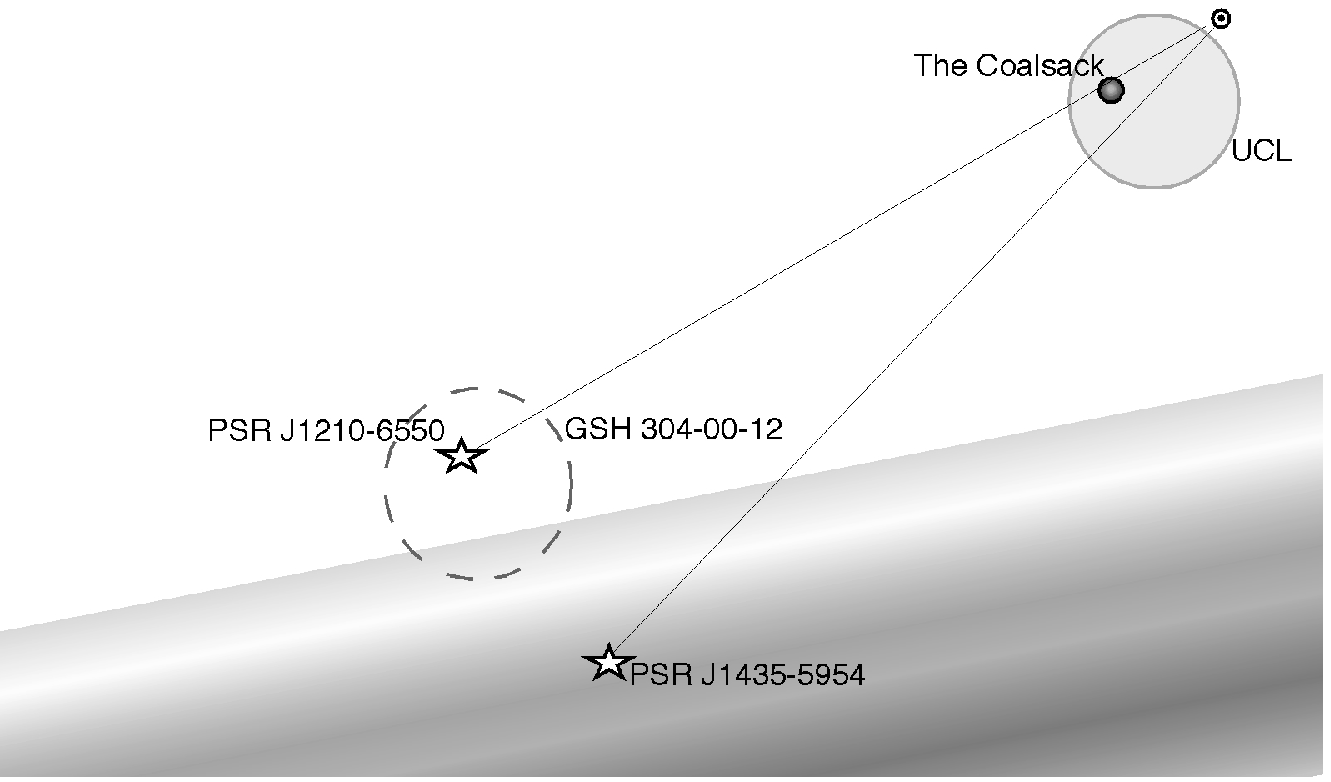}
\caption{A sketch of the relevant Galactic features is shown, with the Sun in the upper right.  The UCL and the Coalsack are located based on the observations in \citet{degeus1992} and \citet{crawford1991}.  The pulsar distances are based on the NE2001 electron density model \citep{cordes2002}.  The gray area in the lower part of the figure represents the Carina spiral arm with a pitch angle of 11$^\circ$ \citep{vallee2005} and a distance, in the direction of the Coalsack, of 1.3 kpc based on the results from \citet{seidensticker1989b}.  The HI shell detected by \citet{mcclure-griffiths2001} is indicated by a dashed circle.}
\label{fig:cartoon}
\end{figure*}

\citet{bga2005} used optical polarimetry and a Chandrasekhar-Fermi (CF) analysis \citep{chandrasekhar1953} to estimate the plane-of-the-sky magnetic field strength in the cloud and found a surprisingly large field of B$_\perp$=93$\pm$23~$\mu$G, but one which is consistent, in equipartition, with the thermal pressure of the X-ray emitting gas seen by \citet{bga2004}.  \citet{lada2004} used the H-band polarimetry of \citet{jones1984} and results from their own C$^{18}$O (J=2-1) observations to perform a CF analysis to estimate the magnetic field strength in Tapia's Globule 2, the darkest of the cores in the cloud.  They estimate a polarization angle dispersion of $\sigma_\theta$=0.66~rad ($\approx$38$^\circ$) and hence derive a plane-of-the-sky magnetic field strength of B$_\perp\approx$24~$\mu$G.  We will argue below, in section \ref{s:tapia}, that separating the polarization data probing Tapia's Globule 2 from the surrounding cloud material, a narrower dispersion of polarization angles is found and hence a higher magnetic field estimate for the globule of B$_\perp\approx$65~$\mu$G, consistent with the optical polarimetry estimate.

\citet{lada2004} also used near-infrared photometry to map the density structure in Tapia's Globule 2.  They detect a ring-like column density enhancement that they interpret as a contracting shell of enhanced space density.  \citet{hennebelle2006} modeled the structure seen by \citet{lada2004} and concluded that it is indeed due to a contracting shell, which they predict will lead to an onset of star formation about 10$^4$ years from now.  They further show that the structure requires an external pressure transient to have passed over the cloud about 1.2$\times10^6$ years ago to trigger the cloud collapse.   They speculate that the triggering event might have been the collision of two cloud fragments, giving rise to Tapia's Globule 2.  Models of cloud compression and collapse for magnetized clouds \citep{li2002} can provide a similar column density pattern, but would then require the magnetic field to be oriented close to the line of sight.  \citet{rathborne2009}, while not able to rule out the model of \citet{hennebelle2006}, prefer a model where the structure originates as a merger of two turbulent flows.

\begin{figure*}[t]
\epsscale{1.50}
\plotone{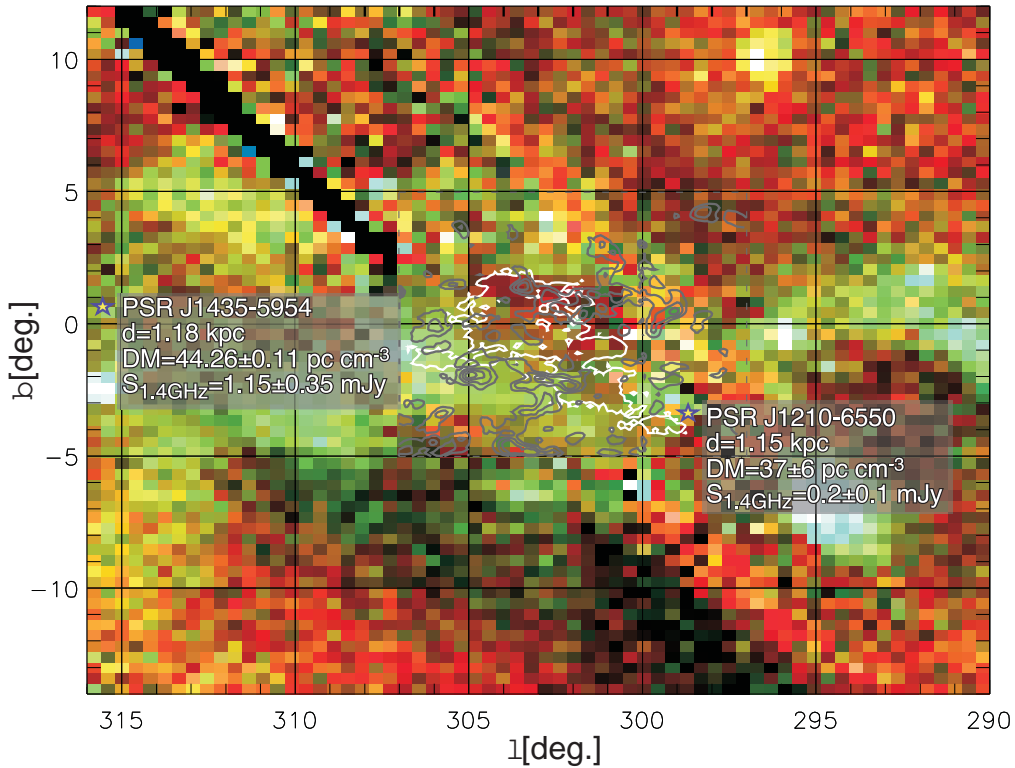}
\caption{Overlaid on a false-colour X-ray map of the Coalsack region (red=1/4~keV, green=3/4~keV, blue=1.5~keV; all ROSAT data) are: (i) the lowest detected CO (J=1-0) contour for the Coalsack (white line; Nyman, Bronfman \& Thaddeus, 1989), and (ii) contours of the polarized 2.4 GHz emission (Duncan et al., 1997; gray contours= 50, 70, 90 and 110 mJ per beam). The 2.4 GHz survey covers $|b|<5^{\circ}$ and we have here cut out data for $297^{\circ}<l<307^{\circ}$, producing a $10^{\circ} \times 10^{\circ}$ area indicated by the dashed gray line. The locations of \psrone and \psrtwo are shown as star symbols. As shown by Andersson et al. (2004), the enhanced X-ray emission surrounding the Coalsack is due to a hot halo enveloping the cloud, which is most likely due 
to an interaction between the cloud and the Upper Centaurus-Lupus super bubble. The polarized radio emission points to a synchrotron origin, consistent with a hot magnetized plasma.}
\label{fig:image}
\end{figure*}

The location of the Coalsack relative to the UCL super bubble, the hot envelope, the strong magnetic field in the cloud and the presence of a possibly contracting core in Tapia's Globule 2, therefore seems to lead to a coherent picture of the Southern Coalsack as a cloud overtaken, compressed and heated by the envelopment into a super bubble and thence as a nearby example of star formation triggered over large distances.  

To further probe this scenario we have acquired Faraday rotation measurements towards the pulsars PSR J1210$-$6550 and PSR JJ1435$-$5954.  The pulsar PSR J1210$-$6550 is located only about 30$\arcmin$ from the lowest $^{12}$CO contour in the South-Westerly-most cloudlet in the map of \citet{nyman1989}.  The pulsar is at an estimated distance of $D$=1.15~kpc, based on the NE2001 electron density model \citep{cordes2002} and hence on the near-side of the Carina arm and should primarily \citep{seidensticker1989b} be probing the material associated with the UCL super bubble and the Coalsack envelope.  For comparison, we also observed the pulsar PSR J1435$-$5954, located about 17$^\circ$ further along the Galactic plane, but at a similar distance of $D$=1.18~kpc.  Figure~\ref{fig:cartoon} shows the sketch of the geometry of the objects.  Of course, it is important to remember the inherent dichotomy between the magnetic field components probed by polarimetry and the CF analysis on the one hand (plane-of-the-sky component) and Faraday rotation measurements (line-of-sight component) on the other.  

Both these pulsars were amongst later discoveries by the Parkes multibeam survey \citep{hobbs2004}. \psrone is a very faint, long-period (4.3 s) pulsar with a very small duty cycle ($\sim$0.1\%), with a period-averaged flux density at 1400 MHz (\Smean) of $\sim$ 0.1 mJy, whereas \psrtwo is a relatively bright pulsar.  
Little is known about their properties apart from what has been deduced from their discovery and initial timing observations. Additionally, we observed \psrten in each observing session, a bright pulsar with a well-known rotation measure (RM), to serve as a `control pulsar' for validating polarimetric calibration and the RM determination. The basic parameters of all three pulsars are listed in Table 1, along with the measurements of RM and the estimates of magnetic field strength derived from our data.  We note that even in the best available pulsar surveys, no further observable targets are available meeting our requirements of 1) on-the-sky proximity to the Coalsack and 2) a distance less than that of the next spiral arm (Carina).

The locations of PSRs J1210$-$6550~and~J1435$-$5954  on the sky and their relation to the known features around the Coalsack region are further illustrated in  Fig.~\ref{fig:image}, which shows an X-ray map in that area (from ROSAT data) along with the detections of CO (J=1-0) from \citet{nyman1989} and the polarized 2.4~GHz emission from \citet{duncan1997}.

\subsection{Magnetic Fields from Rotation Measure Observations}

Rotation measure quantifies the degree of Faraday rotation that electromagnetic waves undergo as they propagate through the ISM from the pulsar to the Earth. This rotation is caused by the interaction of the electromagnetic waves with the magnetized plasma of the ISM and more specifically along the line of sight to the pulsar. For a pulsar located at a distance $D$, the rotation measure (RM) is given by
\be
RM = { e^3 \over 2 \pi m_e^2 c^4 } \int _0 ^D \nele (l) \vecB (l) . d \vecl
\ee
where d\vecl is the path vector element in the direction of wave propagation; $e$ and $m_e$ are the charge and mass of electron, respectively; \nele is the free electron density, \vecB is the magnetic field vector, and $c$ is the speed of light. The above equation 
can be simplified to 
\be
RM = 0.812 \int _0 ^D \nele (l) \vecB (l) . d \vecl 
\ee
where D is in parsecs,  \nele is in per cubic centimeter, \vecB is in microgauss and RM is in radians per square meter. The integral of \nele along the Earth-pulsar line of sight is the dispersion measure (DM), and is given by
\be
DM = \int _0 ^D \nele (l) dl 
\ee
and is usually expressed in units of parsecs per cubic centimeter (\dmu; i.e. D in pc and \nele in \neu). 

%%%%%%%%%%%%%%%%%%%%%%%%%%%%%%%%%%%%%%%%%%%%%%%%%%%%%%%%%%%%%%%%
% Pulsar parameters and measurements 
%%%%%%%%%%%%%%%%%%%%%%%%%%%%%%%%%%%%%%%%%%%%%%%%%%%%%%%%%%%%%%%%
\begin{table*}
\begin{center}
\caption{\label{tab:table}Basic parameters and measured quantities of observed pulsars} 
\vskip 0.25cm
\begin{tabular}{llll}
\hline
\hline
Pulsar 				& \psrten			& \psrone 		& \psrtwo 		\\
\hline
Pulse Period (ms)$^a$		& 198.4514 			& 4237.0102		& 472.9954		\\
Dispersion Measure (\dmu)$^a$	& $116.156 \pm 0.002$		& $37 \pm 6$		& $44.26 \pm 0.11$	\\
Distance (kpc)$^b$ 		& 2.88				& 1.15			& 1.18			\\
Galactic longitude ($^{\circ}$)	& 291.31			& 298.77		& 315.58		\\
Galactic latitude ($^{\circ}$)	& $-$7.13			& $-$3.29		& 0.39			\\
Flux at 1400 MHz \Smean (mJy)	& $2.2 \pm 0.3 $ 		& $0.2 \pm 0.1$	& 	$1.15 \pm 0.35$	\\
Linear polarization (\%)	& 75				& 35 			& 10			\\
Pulse width $W_{50}$ (ms)	& $2.1 \pm 0.1$			& $35 \pm 1$ 		& $18 \pm 1$		\\
Rotation measure (\rmu)		& $-84 \pm 3$			& $-32 \pm 4$		& $-24 \pm 7$		\\
Magnetic field strength
$B_{\parallel}$ ($\mu$G)	& $-0.95 \pm 0.03$		& $-1.1 \pm 0.2$	& $-0.7 \pm 0.2$	\\
\hline
\hline
\end{tabular}
\vskip 0.25cm
$^a$ From the ATNF Pulsar Catalog (also see \citet{hobbs2004}) \\
$^b$ Estimate based on the NE2001 model \citep{cordes2002}
\end{center}
\end{table*} 
%%%%%%%%%%%%%%%%%%%%%%%%%%%%%%%%%%%%%%%%%%%%%%%%%%%%%%%%%%%%%%%%
% End Table
%%%%%%%%%%%%%%%%%%%%%%%%%%%%%%%%%%%%%%%%%%%%%%%%%%%%%%%%%%%%%%%%

The quantity DM can easily be determined from the arrival time delays measured at different frequencies across the observing observing band (e.g. Lorimer \& Kramer 2005) and, consequently, for all catalogued pulsars which are observable in radio this is a well-known quantity. Thus, a measurement of RM enables a direct estimation of the magnetic field strength weighted by the free electron density. The mean value of the line-of-sight component of \vecB is thus given by 
\be
\langle \Bpara \rangle = { \int _0 ^D \nele (l) \, \vecB (l) . d\vecl \over \int _0 ^D \nele (l) \, dl } = 1.232 \, { RM \over DM }  
\ee
The rest of this paper is organized as follows. We briefly describe our observations and data reduction in \S~\ref{s:obs}, determination of rotation measures in \S~\ref{s:rm} and a re-analysis of near infrared polarization data in \S~\ref{s:tapia}. In the later sections (\S~\ref{s:res}~and~\ref{s:disc}) we present our main results and discuss 
their implications for the related physical models. Our conclusions are presented in \S~\ref{s:conc}.
 
\begin{figure}
%\epsscale{.80}
\begin{center}
\includegraphics[height=8.0cm, angle=270]{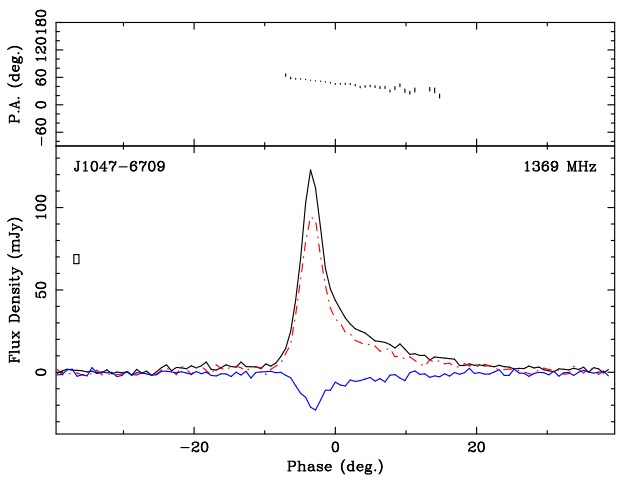}
\end{center}
%\plotone{J1047-6709.eps}
\caption{Polarimetric profiles of \psrten at 1.4 GHz from observations made
with the Parkes multibeam receiver (MJD = 55190). The blue and red lines 
represent circular and linear polarization respectively, while the black lines 
represent the total intensity. Only $80^{\circ}$ of pulse rotation is shown,
and absolute position angles at infinite frequency are shown after correction for 
a rotation measure of $-80$ \rmu.}
\label{fig:psrten}
\end{figure}

\section{Observations and Data Reduction} \label{s:obs}

Pulsar observations were made using the central beam of the 13-beam multibeam receiver \citep{lister1996} on the Parkes 64-metre telescope. This receiver is a dual-channel cryogenic system sensitive to orthogonal linear polarizations with a system equivalent flux density of 36 Jy. 
In all observing sessions, signals centered at 1369 MHz with a bandwidth of 256 MHz were recorded using the Parkes Digital FilterBank 3, which is capable of producing high-resolution profiles in both time and frequency (up to 1024 phase bins and 2048 spectral channels). For all our observations, the full Stokes profiles were recorded across the 256 MHz bandwidth split into 1024 frequency channels, with a 512-bin resolution across the pulse period, over 10-second sub-integrations. The combination of such short sub-integration and high time and frequency resolutions enable a high efficiency in excising the data segments corrupted by radio frequency interference (RFI). 

We collected data through multiple observing sessions conducted between December 2009 and April 2010. The integration times varied depending on the anticipated flux densities of pulsars; \psrone was typically observed for durations of 1--2 hr, whereas \psrtwo for 0.5--1 hr and \psrten for 15--30 min. Due to its low flux density \psrone requires long integrations to enable high quality detections ($\sim$ 1 hr to reach S/N $\sim$ 40).  Each pulsar observation scan was preceded by a short (2-min) observation of the pulsed calibration signal, which is linearly polarised and broadband, and injected into the feed at $45^{\circ}$ to the two signal probes. In addition, we made observations of the radio galaxy Hydra A (3C218), which is assumed to have a flux density of 43.1 Jy at 1.4 GHz, in each observing session in order to enable flux calibration of the data. 

For off-line data processing, we used the {\sc psrchive} pulsar data analysis software package \citep{hotanetal2004}. In brief, the raw data were first subjected to a process of RFI excision, whereby any frequency channels containing strong interference and sub-integrations affected by strong impulsive interference were given zero weight. In addition, the upper and lower 5\% of the band at the band edges were given zero weight due to a low gain and other instrumental effects in these ranges. Using the pulsed calibration signal as reference, variations in instrumental gain and phase across the band were removed and corrected for parallactic angle variations. The data were then summed in time to obtain four integrated Stokes profiles for each frequency channel. A full polarimetric calibration modeling of the receiver as discussed in \citet{vanstraten2004} is not critical for the RM determination and therefore not considered in our analysis. The position angles were then computed for each phase bin from the Stokes Q and U parameters and the flux-density scale was determined from the observations of the Hydra calibration observations. 

\section{Analysis}

\subsection{Determination of Rotation Measures} \label{s:rm}

\begin{figure}[t]
\epsscale{.80}
\begin{center}
%\plottwo{1210.eps}{J1435-5954.eps}
\includegraphics[height=8.0cm, angle=270]{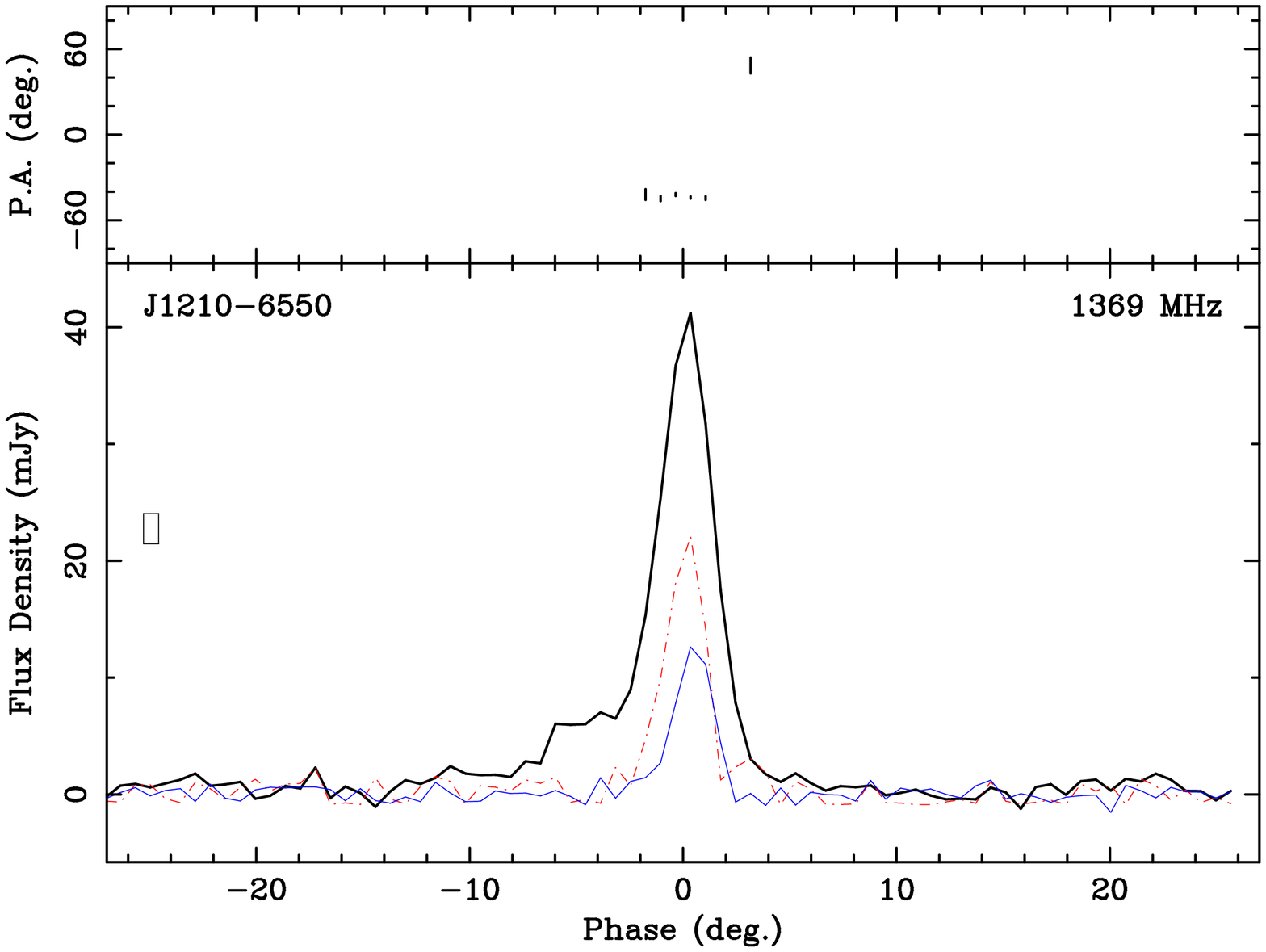}
\vskip 0cm
\includegraphics[height=8.0cm, angle=270]{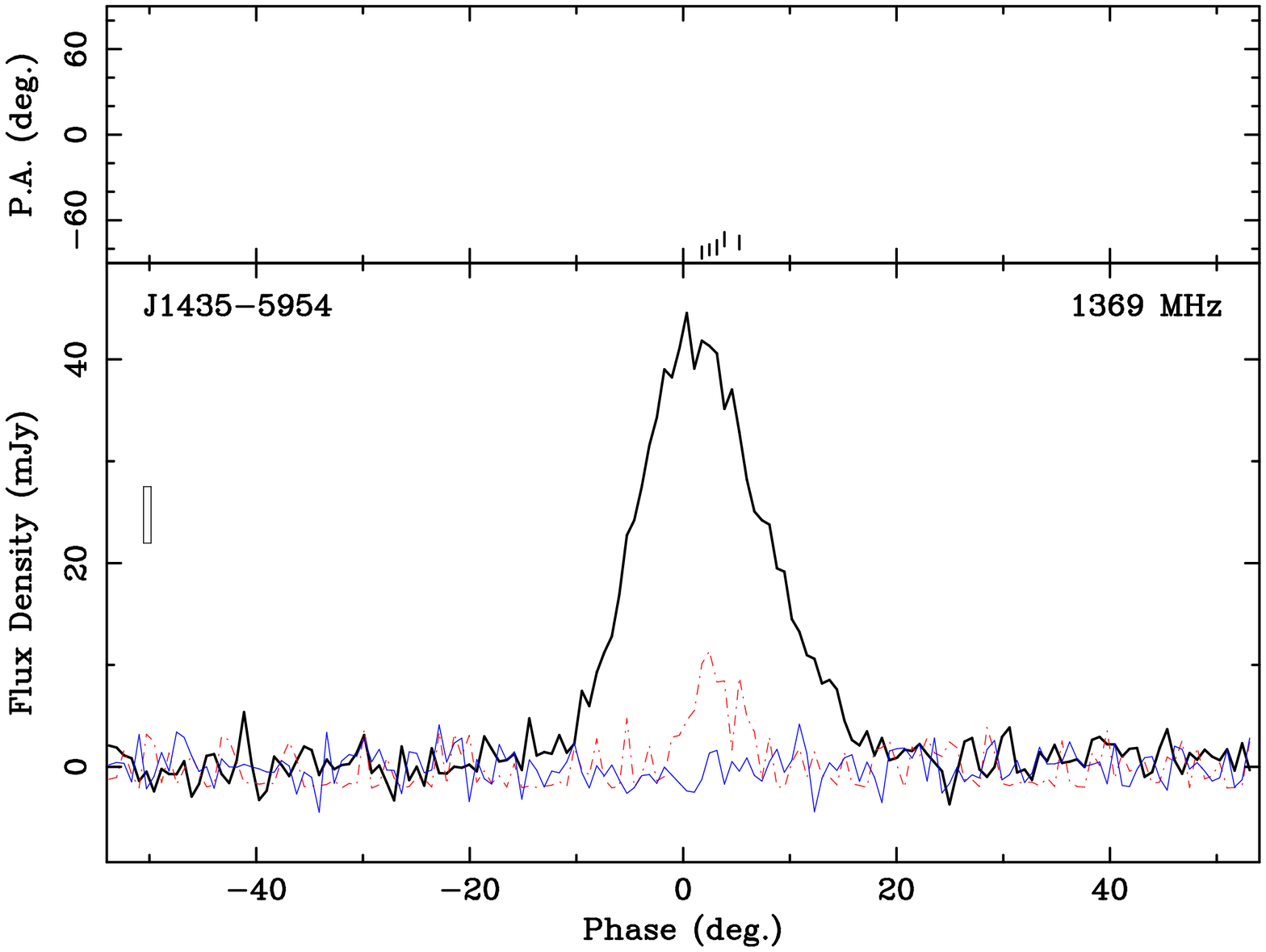}
\end{center}
\caption{Polarimetric profiles of \psrone ({\it top}) and \psrtwo ({\it bottom}) at 1.4 GHz 
from observations made with the Parkes multibeam receiver (MJD = 55190 and 55302 respectively). 
The blue and red lines represent circular and linear polarization respectively, while the black lines 
represent the total intensity. \psrone is moderately polarized, whereas for \psrtwo, the measured 
linear polarization is only $\sim$8\% of the total intensity.}
\label{fig:pulsars}
\end{figure}

Polarimetric profiles obtained from our analysis are shown in Figs.~\ref{fig:psrten}~and~\ref{fig:pulsars}. While \psrone is very faint, it shows a substantial degree of linear polarization ($\sim$35\% of the total intensity). \psrtwo is relatively bright in comparison, however the degree of polarization is much lower (linear and circular polarizations are $\sim$8\% and $\sim$3\% of the total intensity, respectively). 
The signal-to-noise ratios (S/N) of the linearly polarized intensities are therefore not very large and limit the significance achievable in the RM estimates. 
Flux densities of both pulsars show significant variations with time, presumably due to long-term interstellar scintillation effects expected at such moderate DMs 
\citep[e.g.][]{bhat1999}, and the quoted numbers are mean values from multiple measurements made over time spans of several months. 

Our initial estimates of RM were determined using the  standard procedure, where we searched for a peak in the total linearly polarized intensity  $ L = ( Q^2 + U^2 )^{1/2} $ obtained by summing the calibrated data in frequency \citep[e.g.][]{hanetal2006}. This search is carried out over a large range of RM, up to $\pm$ 1000 \rmu, in steps of 1 \rmu. Further, using the RM value corresponding to the peak, we summed the data to form the upper and lower band profiles. These were then used  
to refine the RM estimate by taking the weighted mean of the position-angle differences  between the upper and lower bands across the profile, with the weight inversely proportional to the square of the error in position-angle difference for each pulse phase bin. The RM is then recomputed and the procedure is iterated until convergence. 

In order to confirm and cross-check our initial estimates, as well as to obtain more robust RM measurements, we applied the RM determination method developed by \citet{noutsos2008}, where a quadratic function was fitted to the PA vs frequency across the full observation band. These position angles are computed from the Stokes parameters Q and U that are summed across the phase bins corresponding to the ``on'' pulse.  This quadratic fitting algorithm finds the best fit by means of a Bayesian likelihood test, and uses a fitting function of the form, $ PA = PA_0 + c^2 RM ( 1 / f_j^2 - 1 / f_0^2 ) $, where $ PA_0 $ is the PA at frequency $ f_0 $ and $ f_j $ is the frequency channel $j$. The main advantage of this method is that it accounts for the $ 180^{\circ} $ ambiguity inherent in the computation of the mean PAs from the Stokes Q and U. The free parameters $PA_0$ and RM were stepped through the ranges (0, $\pi$) 
rad and (--1000, 1000) \rmu in $1^{\circ}$ and 1 \rmu, respectively. As demonstrated by \citet{noutsos2008}, this method is more robust and yields more reliable 
RM estimates. The main caveat however is that the signal-to-noise ratio (S/N) per frequency channel or sub-band needs to be sufficiently large for this method to yield meaningful results. 
While we were able to apply this method successfully to both PSRs J1047$-$6709 and J1210$-$6550 (cf. Fig 5), it was not feasible in the case of PSR J1435$-$5954 owing to the very low 
signal-to-noise of its linearly polarised flux (cf. Fig 4). As a result, the RM estimate of this pulsar is derived from the standard method as discussed earlier in this section.

The final estimates of RM derived from our analysis are summarized in Table 1. For the control pulsar \psrten, we obtained four independent measurements over a time span of four months, which are found to be consistent within their 1 $\sigma$ measurement uncertainties, with a mean value of $ -84 \pm 3 $ \rmu. For comparison, values from the published literature are $ -73 \pm 3 $ \rmu \citep{hanetal2006} and $ -79 \pm 2 $ \rmu \citep{noutsos2008}. Thus, our RM estimate is consistent (within 2 $\sigma$) with the more recent measurement of \citet{noutsos2008}, though somewhat larger. 

\begin{figure}[t]
\epsscale{.80}
\begin{center}
%\plottwo{1047-aa16.eps}{1210-aa16.eps}
\includegraphics[height=7.0cm, angle=270]{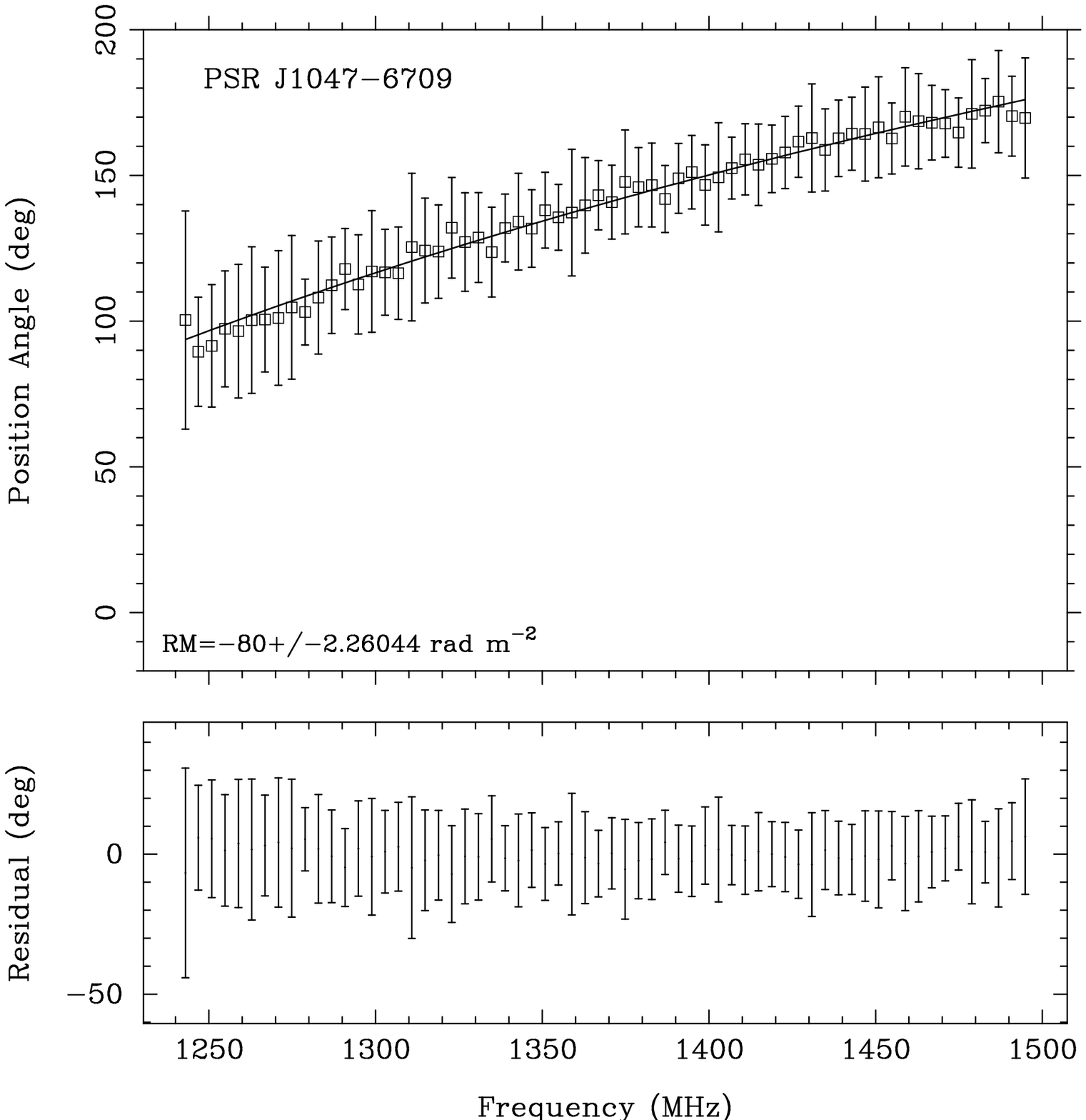}
\vskip 0.25cm
\includegraphics[height=7.0cm, angle=270]{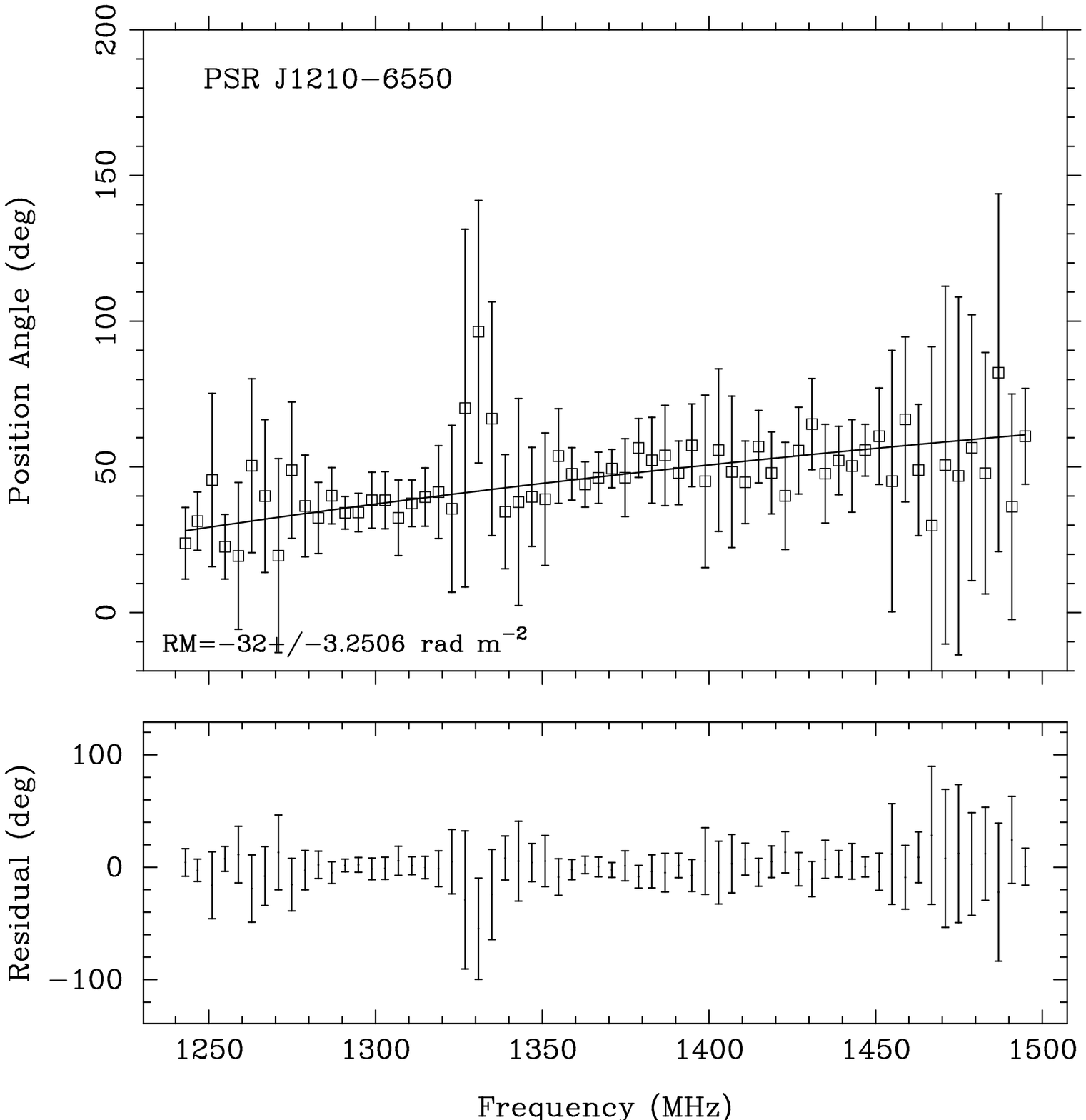}
\end{center}
\caption{Position angle vs frequency plots for \psrten and \psrone, illustrating the RM estimation using the quadratic fit method (see \S~\ref{s:rm} for details). 
The best fit RM determined toward \psrone is $ -32 \pm 4 $ \rmu, which translates to $\Bpara \sim -1.1 \pm0.2$ \muG. These position angles are computed 
from the Stokes Q and U. For \psrone, they are summed over the {\it four} on pulse phase bins that show polarized intensity above a 5-$\sigma$ threshold. 
The large errors on the PAs of this pulsar are thus due to its relatively low degree of polarization and a small number of on pulse phase bins.}
\label{fig:pafit}
\end{figure}

For both \psrone and \psrtwo, we have obtained 2--3 independent measurements (with high significance) over a time span of 3--4 months. The resultant average values are $ - 32 \pm 4 $ \rmu and $ -24 \pm 7 $ \rmu for \psrone and \psrtwo, respectively. The relatively low significance on the RM estimate of the latter is primarily to do with its low degree of polarization 
($\sim$8\%). In comparison, \psrone exhibits a moderately high linear polarization ($\sim$35\%) and hence its RM estimated with a relatively higher significance (Fig.~\ref{fig:pafit}). However, due to its very narrow pulse (duty cycle of $\sim$1\%),  typically only {\it four} phase bins span the on pulse emission with our 512-bin resolution across the pulse period. The consistency between  multiple measurements of each pulsar and the general agreement seen between the published and our measured RM estimates for the control pulsar give us confidence on the reliability of our RM measurements. 

For the current analysis, the dispersion measures towards each of the pulsars under consideration were taken from the ATNF Pulsar Catalog \citep{manchester2005}; for both
\psrone and \psrtwo, the DM measurements ($37 \pm 6$ \dmu and $44.26 \pm 0.011$ \dmu, respectively) are from  \citet{hobbs2004} that reported the original discoveries of these objects. 
 
\subsection{The Magnetic Field in Tapia's Globule 2} \label{s:tapia}

As discussed in the introduction, \citet{lada2004} used near infrared polarimetry from \citet{jones1984} and their own deep $^{18}$CO observations of Tapia's Globule 2 to estimate the magnetic field strength in this dark core.   We here discuss this result in the light of a reanalysis of the polarization data.  \citet{jones1984} note about their H-band polarization map that:

\textit{``Immediately apparent in Figure 4 is a definite change in polarization angle between the Northern block (block E [...]) and the block centered on the globule (block D).   Also the polarization within the 120$\arcsec$ radius circle [the globules ``half-extinction radius"] is about a factor 2 higher than in block E"}. 

\begin{figure}[t]
\epsscale{1.00}
\plotone{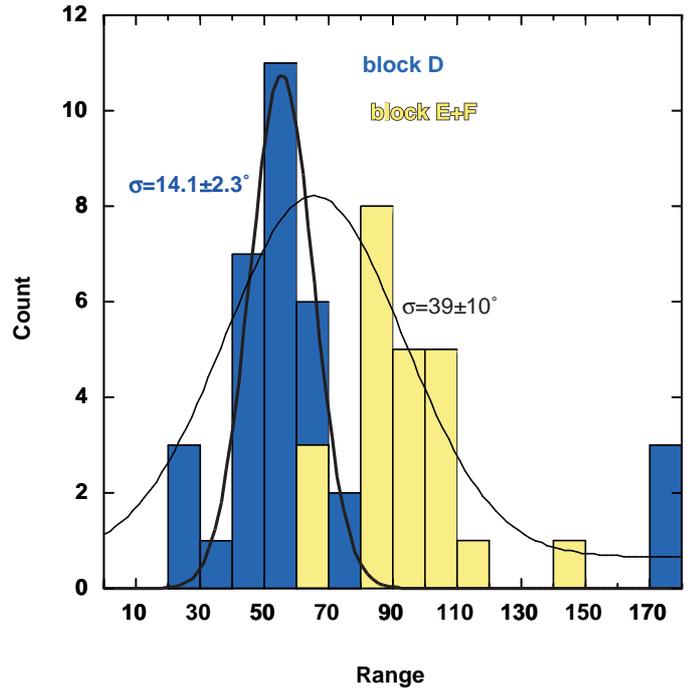}
\caption{The distribution of H-band polarization angles from \citet{jones1984} shows a distinctly bi-modal distribution between the stars behind Tapia's Glubule 2 (``block D") and the surrounding Coalsack material.  For the full sample a position angle dispersion of $\Delta\theta\sim$39$^\circ$ is deduced.  If only the stars in ``block D" are used, an angle dispersion of $\Delta\theta\sim$14$^\circ$ is found implying a magnetic field of B$_\perp\approx$65~$\mu$G.}
\label{fig:pol_disp_fig}
\end{figure}

Figure \ref{fig:pol_disp_fig} shows the distribution of polarization angles from \citet{jones1984} where we have color coded the data from ``block D" in blue (dark grey) and those from blocks E and F in yellow (light grey).  The change in the average value of the polarization angles between the areas is clearly seen.   A student's t-test analysis comparing the polarization angles of the ``block D" sample vs. the combined ``block E" and ``block F" samples yields a t-probability of P=0.0003 and hence shows that the two populations are distinct.  The position angle dispersion for the full sample is $\sigma_\theta$=39$\pm$10$^\circ$.  However, if we follow \citet{jones1984} and separate out only the ``block D" polarization as tracing the field in Tapia's Globule 2, we derive a position angle dispersion of $\sigma_\theta$=14$\pm$2$^\circ$.

\section{Results} \label{s:res}

The estimates of \Bpara derived from our RM measurements are summarized in Table 1. The measured field strength along the sightline toward \psrone is only $-1.1 \pm 0.2$ \muG, which is much 
lower than the plane of the sky component (\Bperp = $93\pm23$ \muG) estimated by \citet{bga2005}. The field strength estimated toward \psrtwo\, is somewhat lower, \Bpara = $-0.7 \pm 0.2$ \muG, but again is directed away from us.  

For our reanalysis of the NIR polarimetry in Tapia's Globule 2, if we retain the values of turbulent velocity ($\sigma_v$=0.23~km~s$^{-1}$) and space density ($n$=10$^4$~cm$^{-3}$) derived by \citet{lada2004}, but use the narrower position angle dispersion, we estimate a plane-of-the-sky magnetic field strength for Tapia's Globule 2 of B$_\perp\approx$\,64\,$\mu$G.  This higher value is consistent (within 1.5$\sigma$) with the field strength (B$_\perp$=93$\pm$23~$\mu$G) derived by \citet{bga2005} using optical polarimetry of the full Coalsack cloud.  It is, however, significantly higher than a value of \Bperp$\approx$\,25\,$\mu$G derived by \citet{lada2004} based on equipartition arguments between magnetic, gravitational and kinetic energies, which is also very close to their CF-analysis based value.  

\citet{rathborne2009} have used a detailed analysis of the C$^{18}$O (J=1-0) emission to argue that the globule is in fact the confluence of two subsonic flows.  They estimate the average space density of the globule to $n\approx$2.7$\times10^3$~cm$^{-3}$.   They, however, also detect CS emission from the inner parts of the globule.  The CS molecule has a high critical density and is expected to trace high-density gas ($n\gtrsim$10$^5$~cm$^{-3}$).  Finally, the spatial resolution of Jones et al's polarimetry data is not high enough to fully sample the magnetic field variations on the scale of the observations of \citet{rathborne2009}.  Hence, a detailed understanding of the magnetic field in the core of the globule will require additional observations.

\section{Discussion} \label{s:disc}

A number of theoretical studies have addressed the characteristics of super bubble evolution \citep{weaver1977,tomisaka1998,stil2009} and of the effects of a magnetized interstellar cloud being overtaken by a supernova remnant (SNR) \citep{maclow1994,melioli2005,leao2009}.  While there are no detailed modeling studies specifically addressing a magnetized cloud enveloped in the relatively slowly expanding super bubble, we can gain some insights into the physics, and estimate the characteristics, of such a cloud by viewing it as a detached part of the inner surface of a super bubble wall or by comparing to the studies of a cloud enveloped by a SNR.  As shown by most super bubble models starting with \citet{weaver1977}, the inside of the super bubble shell will reach temperatures in excess of 10$^6$~K and should reach thermal pressures of P$\sim$10$^6$~K~cm$^{-3}$  \citep{tomisaka1998}.  While the magnetic field is swept up and enhanced, only marginal magnification of the field strength is expected for the inside of the bubble shell \citep{tomisaka1998}.  In contrast, \citet{maclow1994} find that the field comes into equipartition with the gas pressure for the cloud overtaken by a SNR and that the magnetic field protects the cloud from being shredded by the shock.   \citet{leao2009} have calculated the star formation efficiency for clouds enveloped by a SNR driven by an energy injection of 10$^{51}$~erg, but only briefly discuss the results for SNR in the radiative phase, where the expansion velocity is more comparable to that for the UCL super bubble.   Nonetheless, based on their Figure 9, the predicted imminent onset of star formation in the Coalsack \citep{hennebelle2006} is consistent with being triggered by the interaction with the UCL super bubble.

As shown by \citet{tomisaka1998} and \citet{stil2009}, an expanding super bubble will sweep up the ambient magnetic field, compressing it in the direction perpendicular to the undisturbed field direction and leaving an almost field-free bubble interior.   Those calculations show a precipitous drop of the magnetic field at the inside of the bubble wall. Hence, by treating the cloud as a detached part of the wall, we expect that the magnetic field frozen into the cloud will also show an abrupt drop-off.   As noted by \citet{maclow1994}  the initial compression of a magnetized cloud is expected to reverse on time scales short compared to the lifetime of the bubble.  Moreover, in the simulations of super bubbles, the magnetic field leads to relatively thick shells -- particularly in the direction perpendicular to the ambient field \citep{stil2009}.  However, \citet{stil2009} point out that including realistic cooling functions leads to thinner shells and stronger magnetic fields than in adiabatic simulations.  Since the average magnetic field, as traced by the optical polarimetry, runs NE-SW, such corrections would likely be particularly important for the location of \psrone, where cooling flows from the dense cloud likely produces a much more efficient cooling than in the locations where the cloud and surrounding hot gas are connected across the magnetic field lines.

Because the field inside the super bubble in the \citet{stil2009} simulations is weak and turbulent, most of the predicted rotation measure will be due to the shell.  On a large scale, \citet{stil2009} therefore estimate the rotation measure for a 10~Myr old bubble, seen for sightlines perpendicular to the undisturbed field, to be less than $\sim$ 20~rad~m$^{-2}$, dominated by asymmetries in the shell.  For sightlines along the undisturbed field, relatively large rotation measures are seen close to the tangent of the shell while the rotation measure in the projected center of the bubble significantly lower.  These predictions are generally consistent with our observations through the UCL super bubble.  \citet{vallee1983} observed extragalactic background sources through the Gum Nebula and derive an average RM in the inner part of the Gum Nebula of $\sim$130 rad~m$^{-2}$.  However, the Gum Nebula is located at \textit{l} $\sim$270$^{\circ}$ and hence much closer to the projected direction of the Galactic magnetic field.  Also, as \citet{vallee1983} used extragalactic background sources, contributions from the rest of the Galaxy along the line of sight cannot be excluded.  As the simulations by \citet{stil2009} show, for this geometry a fairly large RM is expected.

The line of sight towards PSR J1435$-$5954 is located at a larger Galactic longitude (\textit{l}$\sim$316$^{\circ}$) than that of PSR J1210$-$6550 and hence at a larger angle relative to the nominal undisturbed Galactic magnetic field.   The lower RM seen towards this pulsar, as compared to the one for PSR J1210-6550, is thus also in line with the models by \citet{stil2009}.

Based on the results from the CF analyses from optical \citep{bga2005} and NIR (\citet{jones1984, lada2004}; our Section~\ref{s:tapia}), the results presented here thus seem to indicate that both the sightlines probed fall in the ``diffuse" part of the super bubble, including the one towards PSR J1210$-$6550.  This then suggests that the edge of the magnetized Coalsack cloud is quite sharp and located close to the lowest CO contour -- at least in the direction of the magnetic field.  This is consistent with models, however needs to be tested in detail.  Further observations and models specifically targeted at simulating a magnetized cloud enveloped in a super bubble with parameters tailored to those of the Coalsack and the UCL super bubble are required to address the viability of this interpretation.  

\section{Conclusions} \label{s:conc}

We have acquired rotation measure data for two pulsars behind the Upper Centaurus-Lupus super bubble, one of which probes a sightline very close to the Southern Coalsack.   
While constituting a small sample, these two are the only readily observable pulsars to provide a specific probe of the Faraday rotation in the 3-dimensional vicinity of the cloud. 
We find that the results for both lines of sight are consistent with the predictions of models for a magnetized super bubble without internal clouds.  Since earlier estimates of the magnetic field strength in the Coalsack indicate a strong field (at least in the plane of the sky), this indicates that the magnetized cloud is either very sharply bounded or that the field in the cloud is oriented almost completely 
in the plane of the sky.

We have shown that the observational data set available for the Coalsack is consistent with the hypothesis of a cloud relatively recently overtaken by a super bubble and therefore at the brink of star formation. Specific modeling of the interaction of the Coalsack and the UCL will be required to confirm this picture; the proximity of the cloud and the wealth of observational data are promising to provide important constraints on such models.

\bigskip

\noindent
%\acknowledgments
{\it Acknowledgments:}
We are grateful to Willem van Straten and Aristeidis Noutsos for discussions on polarimetric analysis, Ravi Sankrit and Tim Robishaw for discussions on super-bubble models and interstellar magnetic fields, and Matthew Bailes for his encouragement and support to this project. We thank an anonymous referee who provided useful comments that helped to improve the paper. The Parkes Observatory is part of the Australia Telescope National Facility, which is funded by the Commonwealth of Australia for operation as a National Facility managed by CSIRO.

\bibliography{bgbiblio_tot}

\begin{thebibliography}{30}
\expandafter\ifx\csname natexlab\endcsname\relax\def\natexlab#1{#1}\fi

\bibitem[{{Andersson} {et~al.}(2004){Andersson}, {Knauth}, {Snowden},
  {Shelton}, \& {Wannier}}]{bga2004}
{Andersson}, B.-G., {Knauth}, D.~C., {Snowden}, S.~L., {Shelton}, R.~L., \&
  {Wannier}, P.~G. 2004, \apj, 606, 341

\bibitem[{{Andersson} \& {Potter}(2005)}]{bga2005}
{Andersson}, B.-G. \& {Potter}, S.~B. 2005, \mnras, 356, 1088

\bibitem[Bhat et al.(1999)]{bhat1999} Bhat, N.~D.~R., Rao, 
A.~P., \& Gupta, Y.\ 1999, \apjs, 121, 483 

\bibitem[{{Cambr{\' e}sy}(1999)}]{cambresy1999}
{Cambr{\' e}sy}, L. 1999, \aap, 345, 965

\bibitem[{{Chandrasekhar} \& {Fermi}(1953)}]{chandrasekhar1953}
{Chandrasekhar}, S. \& {Fermi}, E. 1953, \apj, 118, 113

\bibitem[{{Cordes} \& {Lazio}(2002)}]{cordes2002}
{Cordes}, J.~M. \& {Lazio}, T.~J.~W. 2002, ArXiv Astrophysics e-prints

\bibitem[{{Crawford}(1991)}]{crawford1991}
{Crawford}, I.~A. 1991, \aap, 247, 183

\bibitem[{{de Geus}(1992)}]{degeus1992}
{de Geus}, E.~J. 1992, \aap, 262, 258

\bibitem[{{Duncan} {et~al.}(1995){Duncan}, {Stewart}, {Haynes}, \&
  {Jones}}]{duncan1995}
{Duncan}, A.~R., {Stewart}, R.~T., {Haynes}, R.~F., \& {Jones}, K.~L. 1995,
  \mnras, 277, 36

\bibitem[Duncan et al.(1997)]{duncan1997} Duncan, A.~R., Haynes, 
R.~F., Jones, K.~L., \& Stewart, R.~T.\ 1997, \mnras, 291, 279 

\bibitem[{{Franco}(1989)}]{franco1989}
{Franco}, G.~A.~P. 1989, \aap, 215, 119

\bibitem[Han et al.(2006)]{hanetal2006} Han, J.~L., Manchester, 
R.~N., Lyne, A.~G., Qiao, G.~J., \& van Straten, W.\ 2006, \apj, 642, 868 

\bibitem[{{Hennebelle} {et~al.}(2006){Hennebelle}, {Whitworth}, \&
  {Goodwin}}]{hennebelle2006}
{Hennebelle}, P., {Whitworth}, A.~P., \& {Goodwin}, S.~P. 2006, \aap, 451, 141

\bibitem[Hobbs et al.(2004)]{hobbs2004} Hobbs, G., et al.\ 2004, 
\mnras, 352, 1439 

\bibitem[Hotan et al.(2004)]{hotanetal2004} Hotan, A.~W., van Straten, W., 
\& Manchester, R.~N.\ 2004, Publications of the Astronomical Society of Australia, 21, 302 

\bibitem[{{Jones} {et~al.}(1984){Jones}, {Hyland}, \& {Bailey}}]{jones1984}
{Jones}, T.~J., {Hyland}, A.~R., \& {Bailey}, J. 1984, \apj, 282, 675

\bibitem[{{Kato} {et~al.}(1999){Kato}, {Mizuno}, {Asayama}, {Mizuno}, {Ogawa},
  \& {Fukui}}]{kato1999}
{Kato}, S., {Mizuno}, N., {Asayama}, S., {Mizuno}, A., {Ogawa}, H., \& {Fukui},
  Y. 1999, \pasj, 51, 883

\bibitem[{{Knude} \& {Hog}(1998)}]{knude1998}
{Knude}, J. \& {Hog}, E. 1998, \aap, 338, 897

\bibitem[{{Lada} {et~al.}(2004){Lada}, {Huard}, {Crews}, \& {Alves}}]{lada2004}
{Lada}, C.~J., {Huard}, T.~L., {Crews}, L.~J., \& {Alves}, J.~F. 2004, \apj,
  610, 303

\bibitem[{{Le{\~a}o} {et~al.}(2009){Le{\~a}o}, {de Gouveia Dal Pino},
  {Falceta-Gon{\c c}alves}, {Melioli}, \& {Geraissate}}]{leao2009}
{Le{\~a}o}, M.~R.~M., {de Gouveia Dal Pino}, E.~M., {Falceta-Gon{\c c}alves},
  D., {Melioli}, C., \& {Geraissate}, F.~G. 2009, \mnras, 394, 157

\bibitem[{{Li} \& {Nakamura}(2002)}]{li2002}
{Li}, Z. \& {Nakamura}, F. 2002, \apj, 578, 256

\bibitem[Lorimer \& Kramer(2004)]{handbook} Lorimer, D.~R., \& Kramer, M.\ 2004, Handbook of pulsar astronomy, Cambridge University Press

\bibitem[{{Mac Low} {et~al.}(1994){Mac Low}, {McKee}, {Klein}, {Stone}, \&
  {Norman}}]{maclow1994}
{Mac Low}, M., {McKee}, C.~F., {Klein}, R.~I., {Stone}, J.~M., \& {Norman},
  M.~L. 1994, \apj, 433, 757

\bibitem[{{McClure-Griffiths} {et~al.}(2001){McClure-Griffiths}, {Dickey},
  {Gaensler}, \& {Green}}]{mcclure-griffiths2001}
{McClure-Griffiths}, N.~M., {Dickey}, J.~M., {Gaensler}, B.~M., \& {Green},
  A.~J. 2001, \apj, 562, 424

\bibitem[Manchester et al.(2005)]{manchester2005} Manchester, R.~N., 
Hobbs, G.~B., Teoh, A., \& Hobbs, M.\ 2005, \aj, 129, 1993 

\bibitem[{{Melioli} {et~al.}(2005){Melioli}, {de Gouveia dal Pino}, \&
  {Raga}}]{melioli2005}
{Melioli}, C., {de Gouveia dal Pino}, E.~M., \& {Raga}, A. 2005, \aap, 443, 495

\bibitem[Noutsos et al.(2008)]{noutsos2008} Noutsos, A., Johnston, 
S., Kramer, M., \& Karastergiou, A.\ 2008, \mnras, 386, 1881 

\bibitem[{{Nyman}(2008)}]{nyman2008}
{Nyman}, L. 2008, {The Southern Coalsack}, ed. {Reipurth, B.}, 222

\bibitem[{{Nyman} {et~al.}(1989){Nyman}, {Bronfman}, \& {Thaddeus}}]{nyman1989}
{Nyman}, L.-A., {Bronfman}, L., \& {Thaddeus}, P. 1989, \aap, 216, 185

\bibitem[Rathborne et al.(2009)]{rathborne2009} Rathborne, J.~M., 
Lada, C.~J., Walsh, W., Saul, M., \& Butner, H.~M.\ 2009, \apj, 690, 1659 

\bibitem[{{Seidensticker} \& {Schmidt-Kaler}(1989)}]{seidensticker1989b}
{Seidensticker}, K.~J. \& {Schmidt-Kaler}, T. 1989, \aap, 225, 192

\bibitem[{{Shapiro} \& {Moore}(1976)}]{shapiro1976}
{Shapiro}, P.~R. \& {Moore}, R.~T. 1976, \apj, 207, 460

\bibitem[Staveley-Smith et al.(1996)]{lister1996} Staveley-Smith, L., et al.\ 1996, 
Publications of the Astronomical Society of Australia, 13, 243 

\bibitem[{{Stil} {et~al.}(2009){Stil}, {Wityk}, {Ouyed}, \&
  {Taylor}}]{stil2009}
{Stil}, J., {Wityk}, N., {Ouyed}, R., \& {Taylor}, A.~R. 2009, \apj, 701, 330

\bibitem[{{Straizys} {et~al.}(1994){Straizys}, {Claria}, {Piatti}, \&
  {Kaslauskas}}]{straizys1994}
{Straizys}, V., {Claria}, J.~J., {Piatti}, A.~E., \& {Kaslauskas}, A. 1994,
  Baltic Astronomy, 3, 199

\bibitem[{{Tomisaka}(1998)}]{tomisaka1998}
{Tomisaka}, K. 1998, \mnras, 298, 797

\bibitem[{{Vall{\'e}e}(2005)}]{vallee2005}
{Vall{\'e}e}, J.~P. 2005, \aj, 130, 569

\bibitem[{{Vallee} \& {Bignell}(1983)}]{vallee1983}
{Vallee}, J.~P. \& {Bignell}, R.~C. 1983, \apj, 272, 131

\bibitem[van Straten(2004)]{vanstraten2004} van Straten, W.\ 2004, 
\apjs, 152, 129 

\bibitem[{{Walker} \& {Zealey}(1998)}]{walker1998}
{Walker}, A. \& {Zealey}, W.~J. 1998, Publications of the Astronomical Society
  of Australia, 15, 79

\bibitem[{{Weaver} {et~al.}(1977){Weaver}, {McCray}, {Castor}, {Shapiro}, \&
  {Moore}}]{weaver1977}
{Weaver}, R., {McCray}, R., {Castor}, J., {Shapiro}, P., \& {Moore}, R. 1977,
  \apj, 218, 377

\end{thebibliography}
\bibliographystyle{apj}

%\begin{thebibliography}{}
%\end{thebibliography}

\end{document}